# Static Detection of Post-Quantum Cryptographic Algorithms in Stripped Binaries for Digital Forensic Examination and Migration Assurance

Muhammad Shaheer Bin Junaid
Independent Researcher
Karachi, Pakistan
*contact@mshaheerbjunaid.com*

**Abstract—** Identifying which post-quantum algorithm a compiled binary implements is a core problem in binary analysis, and it becomes acute once the binary has been stripped, statically linked, and optimised, since the symbols, library dependencies, and runtime behaviour that conventional discovery tools rely on are then gone. This paper presents Kestrel, a static analysis method that identifies an algorithm from the number-theoretic transform constant tables its arithmetic depends on, and applies it to the standardised lattice schemes ML-KEM and ML-DSA, for which no prior method could confirm, from a shipped binary alone, that a quantum-vulnerable algorithm had been replaced by its approved successor. The fingerprints Kestrel derives from public scheme parameters are localised by a normalisation-and-multiset-matching procedure; the false-positive probability is established analytically. Across four independent implementation lineages and all build transformations, including compiler-level obfuscation, Kestrel achieved recall of 128 of 128 with zero false positives. Applied to 6,224 binaries on a production Linux system, it disclosed twelve uncatalogued programs containing ML-KEM, among them the OpenSSH key-exchange program and the container-management stack, where post-quantum code had entered production through the language runtime without the awareness of the projects distributing them. Kestrel distinguishes genuine implementations from advertised claims not backed by the underlying code, attributes each detection to its originating codebase, and, in a forensic disk-image trial, recovered a detection from unallocated space after the deleted binary could no longer be reconstructed. It thus provides a practical basis for cryptographic migration assurance, compliance verification, software supply-chain inspection, and post-quantum forensic examination.

## I. Introduction

The migration to post-quantum cryptography depends on a preliminary step whose difficulty is easily underestimated. Before an organisation can replace the algorithms that a quantum computer would defeat, it must first locate every instance of those algorithms across software accumulated over decades of development. The principle underlying this requirement, restated throughout current migration guidance, is that cryptography which cannot be observed cannot be replaced. A dedicated category of tooling, the cryptographic discovery scanner, exists to produce the necessary inventory, and national standards bodies now treat automated discovery as the foundation of the entire transition.

These tools are, without exception, oriented in a single direction. They identify RSA, elliptic-curve schemes, and Diffie-Hellman, since these are the primitives rendered insecure by Shor's algorithm [4] and therefore the ones a migration programme must remove. That orientation was appropriate while the task was removal, but it leaves a blind spot that is only now becoming apparent. When such a scanner is asked the converse question, namely whether a given binary contains ML-KEM or ML-DSA, it returns no answer, because no fingerprints exist for algorithms standardised only in 2024. The discipline has concentrated on identifying what must be withdrawn, and no instrument confirms what has actually been deployed in its place.

This deficiency ceased to be a purely theoretical concern in 2026. A ransomware family designated Kyber was documented [19] whose Windows variant genuinely encapsulates its AES-256 file key with ML-KEM-1024 in combination with X25519, while a related variant targeting virtualisation hosts claims post-quantum protection but in fact relies on classical RSA. Post-quantum cryptography has thus become an offensive capability in deployed malware, and an examiner confronted with such a sample possesses no established means to identify the primitive, document it in a report, or distinguish a genuine implementation from a fraudulent claim.

The same period produced a corresponding regulatory obligation. Executive Order 14412, Securing the Nation Against Advanced Cryptographic Attacks, signed on 22 June 2026 [18], directs federal agencies to transition their high-value and high-impact systems to post-quantum cryptography, employing it for key establishment by 31 December 2030 and for digital signatures by 31 December 2031, and requires the acquisition regulations to oblige covered contractors to conform to the post-quantum standards on the same schedule. A migration subject to a statutory deadline gives rise to a verification problem that current practice cannot resolve. When a vendor delivers a binary together with an assurance that it has migrated to ML-KEM, no established procedure confirms that claim against the artefact itself rather than against the vendor's declaration. Source-level scans reveal intended use rather than what was shipped; network scans observe traffic rather than the implementation within the file; configuration scans record settings rather than code. As one binary-analysis vendor states in its own technical material, the common approaches to cryptographic inventory each leave the compiled binary itself unexamined [29].

This paper resolves that gap for the lattice schemes that predominate in current deployment, through a method implemented in a tool named Kestrel. The principle on which it rests is well established, whereas its application is novel. ML-KEM and ML-DSA perform their polynomial arithmetic [1], [2] by means of the number-theoretic transform, which requires a table of precomputed constants, namely the powers of a fixed root of unity modulo a fixed prime. These constants are not incidental to the computation; the algorithm produces incorrect results in their absence, so they cannot be removed, and because they are stored as data rather than as executable code, they persist through the transformations that eliminate every signal a name-based scanner depends upon. Compiler optimisation rewrites instructions while leaving constant tables unaltered; symbol stripping removes names while leaving data intact; static linking incorporates the table into the binary rather than excluding it. The function name may be removed, whereas the constant 3329 cannot. To the best of our knowledge, no prior method identifies a post-quantum primitive from these constants within a stripped, statically linked binary, since existing work relies on symbols, library dependencies, runtime behaviour, or network traffic, none of which survives the conditions Kestrel is designed to address. Kestrel is released as open-source software, with its generator, detector, corpus builder, and obfuscation pass provided in full [30].

This paper makes four contributions. First, it defines a derivation procedure that generates the complete fingerprint set for any transform-based lattice scheme from its public parameters alone, emitting every representation in which shipped implementations store the constants, so that coverage does not depend on the manual enumeration of implementations. Second, it presents a detector built upon that procedure, employing a normalisation step and multiset scoring that accommodate the reordered and Montgomery-form layouts produced by optimised builds, and whose false-positive rate can be stated as a quantified value rather than merely asserted. Third, it establishes an implementation-lineage attribution result, confirmed by direct inspection of source code, demonstrating that the number of constant tables a binary carries separates the principal deployed lineages at a granularity that a forensic examiner can defend, which is precisely the form of provenance an investigation requires. Fourth, it articulates the two operational settings the method serves, forensic examination and migration assurance, together with a candid account of its capabilities and limitations. The principle is not specific to post-quantum schemes. Any algorithm whose correctness depends on stored, invariant constants leaves the same kind of fingerprint, so constant-table detection is a general technique for identifying algorithms in stripped binaries.

The paper does not claim that constant matching is itself a new technique. Signature-based tools have located the AES substitution box by its fixed bytes for two decades. The novelty is more specific and defensible: the construction and validation of the first fingerprint set and detector for post-quantum primitives in stripped binaries, at the point at which such primitives have appeared in malware and a federally mandated migration deadline has rendered their verification a matter of compliance. The technique is long established; its application to this problem is not.

The evaluation confirms this on software the method had not previously encountered. Kestrel identifies ML-KEM and ML-DSA at full table match across four independent toolchains and every ordinary build transformation, including symbol stripping, static linking, and a genuine obfuscation pass. A sweep of 6,224 binaries from a production system produced no false positives and, in the same pass, disclosed twelve distinct uncatalogued programs containing post-quantum cryptography, among them a live OpenSSH key exchange and the container-management stack, in which ML-KEM had entered production through the language runtime without the awareness of the projects distributing it. The detector distinguished a genuine implementation from one that merely advertises post-quantum protection, attributed each finding to its source lineage by table configuration alone, and, in a forensic disk-image trial, recovered a detection from unallocated space after the deleted binary had itself become unrecoverable. In contrast to the classical primitives that preceded it, post-quantum cryptography leaves a fingerprint that ordinary compilation cannot remove.

## II. Background

### A. Why the transition looks the way it does

A sufficiently large quantum computer executing Shor's algorithm solves the integer-factorisation and discrete-logarithm problems [4] on which RSA, ECDSA, and Diffie-Hellman depend, which does not merely weaken these schemes but renders them insecure. Symmetric primitives are affected differently; Grover's algorithm reduces the effective key length by half [5], so that AES-256 retains an adequate security margin. The resulting migration therefore concerns public-key cryptography specifically, and it is accompanied by a harvest-now-decrypt-later exposure in which an adversary records ciphertext at present in order to decrypt it once such a machine becomes available. Data whose confidentiality must persist beyond that point is, in effect, already at risk. In 2024 the United States standards body finalised three replacements: ML-KEM for key encapsulation, ML-DSA for digital signatures, and the hash-based SLH-DSA [3]. The first two are lattice schemes and constitute the subject of this work.

### B. Module lattices and the number-theoretic transform

ML-KEM and ML-DSA operate on polynomials of degree 255 whose coefficients lie in the ring of integers modulo a prime q. Their central operation is polynomial multiplication, which, performed directly, incurs a cost on the order of n-squared coefficient multiplications for degree n and is prohibitive at the rate at which these schemes execute. The number-theoretic transform renders the operation practical [8]. It is a discrete Fourier transform performed over a finite field rather than over the complex numbers, and it converts a polynomial into a point-value representation in which multiplication becomes coefficient-wise, thereby reducing the cost to order n log n. The transform is not an optional optimisation; it is prescribed in the standard's own pseudocode, and every conforming implementation performs it.

The transform requires a primitive root of unity. For a modulus q and transform length n, this is an element zeta whose successive powers, reduced modulo q, traverse a complete set of positions before returning to unity. ML-KEM adopts q = 3329 and the root zeta = 17, whereas ML-DSA adopts q = 8380417 and a correspondingly larger transform. The forward transform repeatedly multiplies coefficients by powers of zeta arranged in a fixed, bit-reversed order, and, rather than recomputing these powers at each execution, every implementation precomputes them once into a hardcoded array [6], [7]. It is this array, termed the zeta table, that the present method detects.

The powers are generated by the relation in equation 1, where $\mathrm{BitRev}_7(i)$ denotes the seven-bit reversal of the index i for the length-128 ML-KEM table, and the whole quantity is reduced modulo q:

$$\zeta_{\mathrm{table}}[i] = \zeta^{BitRev7(i)} \bmod q \qquad (1)$$

Reference implementations do not store the plain residues of equation 1. Instead they store the constants in Montgomery form, premultiplied by a factor $R = 2^{16} \bmod q$ so that modular multiplication may be performed efficiently, as expressed in equation 2. This representation is significant for detection, since the same mathematical table appears on disk as an entirely different byte sequence according to the representation an implementation has adopted:

$$\zeta_{\mathrm{mont}}[i] = \zeta^{BitRev7(i)} \cdot R \bmod q, \quad R = 2^{16} \bmod q \qquad (2)$$

The modular reduction performed after each Montgomery multiplication employs a further fixed constant, the value q inverse modulo $2^{16}$, equal to 62209 for ML-KEM. Unlike the table, this value is embedded in the instruction stream as an immediate operand rather than stored as data, a distinction that determines which signals survive stripping and which do not, and which is developed further in Section VI.

### C. Why the constants survive

A compiled binary separates code from data. Executable instructions occupy one region, conventionally the text section, while read-only constants occupy another, conventionally the read-only data section. The two behave very differently under the transformations that software routinely undergoes between authoring and distribution. Compiler optimisation reorders, inlines, and rewrites instructions, but it cannot alter the numeric value of a constant table without changing the program's output, so the table passes through unaffected. Symbol stripping, applied to essentially every production binary because names serve only for debugging, removes the symbol table and with it the function names a scanner would match, yet it does not affect the data section. Static linking, common in contemporary Go and Rust binaries and in embedded firmware, incorporates a library's code and data into the executable, and therefore includes the zeta table rather than excluding it. The governing principle is that the algorithm cannot operate without leaving this fingerprint, and the fingerprint cannot therefore be removed by any transformation that preserves the algorithm. The stability of these constants is independently corroborated in the fault-injection literature, in which the twiddle factors of the transform are studied [28] precisely because they are fixed, security-critical values that a correct implementation must store and apply without alteration. Figure 1 illustrates this directly: the code section is rewritten and the symbol table removed, while the constant table on disk remains byte-for-byte identical.

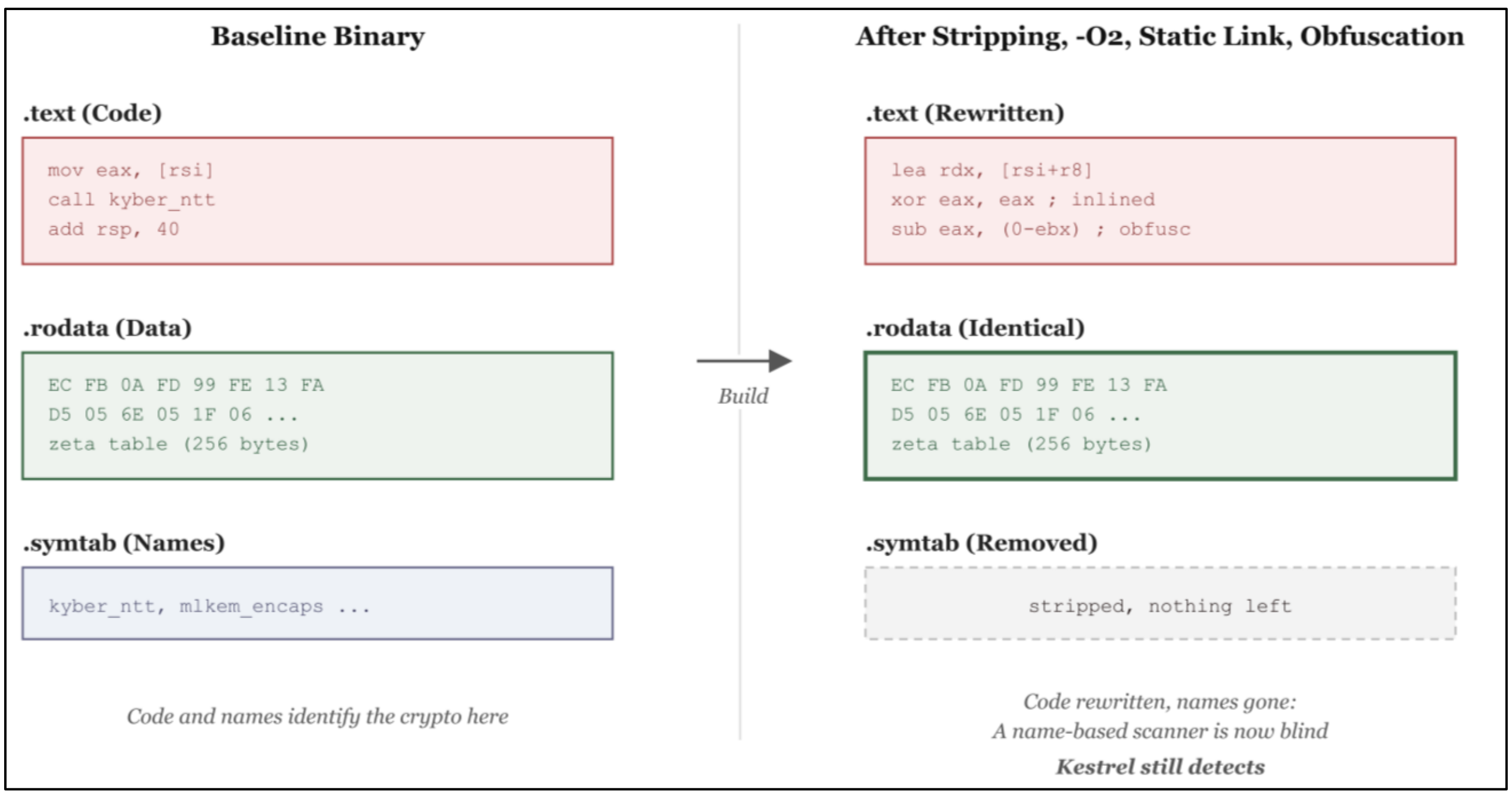

*Fig. 1. Optimisation and obfuscation rewrite .text and stripping erases .symtab,*

## III. Related Work and the Gap

Prior work relevant to this paper falls into three categories, none of which identifies a post-quantum primitive from the static constants of a stripped binary. Figure 2 summarises this comparison and situates the present method against existing approaches.

### A. Cryptographic identification in binaries

The recognition of cryptography in compiled code has been studied for approximately two decades, and it has consistently had to contend with obfuscation. The earliest tools match fixed constants: FindCrypt and Signsrch locate the AES substitution box and standard initialisation values directly within the byte stream, and findcrypt-yara performs the same function through YARA rules. Subsequent work advanced toward semantic identification, recognising primitives by data-flow structure [12], [16], by symbolic execution of candidate loops [13], or by bit-precise emulation that remains reliable under obfuscation [14], [15]. All of this work targets classical primitives, and none provides a signature for a lattice scheme.

The most closely related tool warrants precise description, since it defines the gap through the scope of its own coverage. The open-source crypto-detection project by Kiona [22] constructs YARA rules for stripped, statically linked, and proprietary binaries, which is the surface this paper addresses, and it does so using the same class of technique, namely constant and instruction signatures. Its supported algorithms are ChaCha20, RC4, AES, DES, MD5, SHA-1, SHA-2, SHA-3, and Base64. Post-quantum cryptography appears only once in the repository, within the future-work list, recorded as a possibility to be addressed should the need arise. The apparatus capable of detecting a lattice scheme from its constants has therefore been constructed and applied to every symmetric and hash primitive of consequence, but never to ML-KEM or ML-DSA. This paper extends that established constant-matching approach to the post-quantum setting rather than devising a new one, and it states that inheritance explicitly.

### B. Post-quantum-specific discovery

A smaller and more recent body of work addresses post-quantum algorithms directly, and the most closely related of these substantiates the motivation for the present paper. QED identifies quantum-vulnerable cryptography in software executables [9] and elects to rely on library API names because, by its own account, static features such as look-up tables may be disturbed when a compiler optimises aggressively, whereas an API name persists. QED further states explicitly that it does not address statically linked or self-implemented cryptography, which is precisely the surface examined here. This reasoning is sound for the problem QED defines, and it also identifies the aperture in which the present work operates. The NTT constant tables of ML-KEM and ML-DSA are not disturbed by optimisation, because they constitute read-only data rather than code, and the feature QED sets aside as fragile is therefore, for lattice schemes specifically, the most durable signal available. Where QED reads a name that stripping removes, the present method reads a table that stripping cannot affect.

The QED line of work has continued to develop through 2026, producing lighter, machine-learning-based successors that reduce the false-positive rate [26], [27] of the same category of analysis. These tools remain founded on function symbols, library dependencies, and binary-level attributes rather than on cryptographic constants, and they inherit the same limitation: a stripped, statically linked binary that exposes no names and links no external libraries presents none of the features on which they depend. The present method is complementary rather than competing, since it recovers precisely the signal that these tools forfeit once the symbols have been removed.

A second line of work, by Mallick and colleagues [10], classifies post-quantum implementations dynamically, through the CPU-cycle and memory footprint of a running process and through the key sizes observable in network traffic, thereby distinguishing libraries and identifying post-quantum TLS across a large sample of domains. Earlier versions of that work were framed in terms of fingerprinting, and the distinction from the present paper should be stated explicitly: their signal is runtime and network behaviour, whereas the signal employed here is the static byte content of the stored constant table. Their method requires either a live process or captured traffic, and its accuracy in distinguishing signature schemes reaches approximately the mid-eighties of a percent, because runtime footprints are difficult to separate. A static file recovered from seized media affords neither a process nor traffic, and a constant match, where the table is present, constitutes a near-certain identification rather than a probabilistic estimate. A further recent contribution validates post-quantum public keys that have already been identified [11], verifying their correctness, which presupposes the identification step that this paper provides.

### C. Cryptographic bill of materials tooling

The commercial discovery market expanded rapidly during 2025 and 2026, and several platforms now include binary scanning among their stated methods, so the assertion that no tool examines binaries would be inaccurate and is not the claim advanced here. IBM Quantum Safe Explorer performs static analysis of source and object code [23] and scans compiled binaries, and SandboxAQ, Keyfactor, and Encryption Consulting offer comparable discovery products. The pertinent distinction concerns what these tools scan binaries in order to find. The documented purpose of Explorer with respect to object code is to identify quantum-vulnerable algorithms and hard-coded keys, namely the cryptography that a migration is intended to remove. The market as a whole is characterised by its own analysts as discovery-first, designed to locate and catalogue vulnerable assets. Binarly holds a patent for the generation of a cryptographic bill of materials [29] from binary executables, which confirms that the industry recognises the problem, but a bill-of-materials generation workflow is distinct from the identification of a specific post-quantum primitive from its arithmetic constants. None of these platforms documents that capability for a binary that carries no symbols and executes no process.

The most explicit statement of the gap appears in a migration-planning framework for embedded and defence systems, which begins from the same premise as the present paper and reaches the opposite conclusion. It identifies static linking, stripped binaries, and proprietary implementations as the barriers that defeat discovery, and it observes that, once symbols have been removed, identification reduces to matching known cryptographic constants. It then sets that approach aside, on the grounds that constant matching exhibits high false-positive rates and cannot accommodate custom code, and recommends vendor-supplied documentation as the primary discovery mechanism in its place. That judgement constitutes the motivation for the present work. The constant-matching approach was abandoned because its false-positive behaviour was regarded as unmanageable; Section VI constructs the detector, and Section VIII derives that behaviour as an analytical bound and measures it at zero across 6,224 real binaries, establishing the detector that the framework concluded could not be relied upon.

| | Static, no execution | Reads the binary itself | Detects quantum-safe | Survives strip and static link |
|---|---|---|---|---|
| Name-based scanners (FindCrypt, CBOM) | ✓ | ✓ | ✗ | ✗ |
| Dynamic fingerprinting (2025) | ✗ | ✗ | ✓ | ✗ |
| Network / traffic analysis | ✗ | ✗ | ✓ | ✗ |
| QED (2024) | ✓ | ✓ | ✗ | ✗ |
| **Kestrel** | ✓ | ✓ | ✓ | ✓ |

Fig. 2. Where existing approaches go blind. Kestrel is the only method that reads the binary statically, detects quantum-safe algorithms, and survives the stripping and static linking that erase name-based signals.

## IV. Threat Model and Problem Definition

The method is defined for a single, realistic setting in which an analyst possesses a compiled binary and must determine which cryptographic primitives it implements, without access to source code, without executing it, and without observing it on a network. Two operational instances of this setting motivate the work, and they differ in the identity of the analyst and in the cost associated with an error.

### A. Forensic examination

A device is seized and imaged. The examiner has access to files rather than to processes; a static disk image cannot be executed, and in an incident-response or evidentiary context the execution of recovered malware is frequently inadmissible or unsafe in any event. The examiner must state, in a report that may be subjected to scrutiny in court, what cryptography the sample employs. In this setting the consequential error is the false positive: to assert that a primitive is present when it is not overstates the evidence and constitutes a serious professional failure. The distinction between a genuine and a fraudulent implementation observed in the Kyber ransomware family exemplifies this problem precisely. One variant genuinely incorporates ML-KEM-1024, whereas another claims post-quantum protection without providing it. Only a method that inspects the implementation itself, rather than its self-description, can separate the two, since the genuine variant carries the zeta table and the fraudulent one cannot.

**B. Migration assurance**

A vendor delivers a binary and asserts compliance with a post-quantum requirement, and an auditor must verify that assertion against the artefact itself. In this setting the consequential error is likewise the false positive, although it manifests differently: to report cryptography that is in fact absent permits a non-compliant binary to be certified as migrated, which constitutes a security failure bearing the auditor's signature. A false negative, by contrast, which is a failure to confirm a claim that happens to be true, merely occasions further scrutiny and is therefore costly rather than dangerous. The asymmetry is identical in both instances and governs the detector's design: the scoring is calibrated so that a positive identification is trustworthy, and missed detections are accepted as the less costly error.

Two boundaries are established at the outset and stated explicitly so that they are not misconstrued as omissions. The method addresses lattice schemes that store a transform table, namely ML-KEM and ML-DSA. It does not address SLH-DSA, which is hash-based and stores no such table, nor the FFT-based signature scheme, whose floating-point structure presents a distinct fingerprinting problem, nor the code-based schemes selected for subsequent standardisation. The method further assumes an adversary bounded by ordinary build practice. An implementer who deliberately computes the transform constants at runtime, at a cost of approximately one hundred modular multiplications during initialisation, defeats static detection entirely. This form of evasion is genuine and is discussed in Section X; it is, however, presently absent from every shipped implementation and from the malware observed to date, for the same reason that most software does not strip its constants, namely that no incentive exists to do so.

## V. Fingerprint Derivation

The detector's fingerprints are not composed manually. They are generated from the public parameters of a scheme by a short, deterministic procedure, which constitutes the first of the paper's contributions. A manually compiled signature list is a fixed dataset, covering only those implementations its author has examined. A derivation procedure, by contrast, is a general method: it covers any implementation of the standard, including those not yet written, and it extends to a new scheme through the addition of a small set of parameters rather than through a separate research effort.

The procedure accepts four inputs: the modulus q, the root of unity zeta, the transform length n, and the bit-reversal width w. An initial and easily overlooked point is that a scheme does not store a single table. Depending on the manner in which its transform is written, an implementation may store the forward twiddles, a separate inverse-transform table, and a distinct set of base-case multiplication constants. The generator therefore emits three exponent families, presented in equations 3 and 4: the forward twiddles, the inverse twiddles, and the base-case or mod-root constants.

$$\text{forward: } \zeta^{BitRev7(\mathrm{i})} \bmod q \qquad \text{inverse: } \zeta^{-BitRev7(\mathrm{i})} \bmod q \qquad (3)$$

$$\text{mod-roots: } \zeta^{2\cdot BitRev7(\mathrm{i})+1} \bmod q \qquad (4)$$

These families are not approximate copies of one another. The forward and inverse tables of ML-KEM share exactly one member out of 128, so that a detector scoring only the forward table would identify essentially nothing in a binary that stores its inverse table separately. The forward and mod-root families overlap in approximately half of their members. The consequence is concrete and was the source of a genuine detection gap in an earlier version of this design: an implementation that retains its inverse table as a distinct array is invisible to a forward-only scorer, and the detector must therefore carry and score all three families independently.

Each family is subsequently emitted in the byte-level representations that shipped implementations store, since the same residues appear on disk as different byte sequences. The reference C lineage stores the forward twiddles in Montgomery form as signed sixteen-bit integers, beginning minus 1044, minus 758, minus 359, minus 1517; in little-endian order, minus 1044 corresponds to the bytes EC FB. Python-family ports store the same residues as unsigned sixteen-bit values, beginning 2285, 2571, 2970, 1812, in which the first constant corresponds to the bytes ED 08 and shares no byte with the signed form of the identical value. Implementations written from the specification store plain residues, beginning 1, 1729, 2580, 3289. Vectorised builds store the same residues in a reordered and lane-duplicated arrangement. Because the scorer normalises before matching, as described in Section VI, these representations converge upon a single target set rather than requiring four separate searches; the generator emits them so that the corpus can be indexed and the byte-level search patterns required for the initial inspection can be produced.

A single verification anchors the entire procedure. The first Montgomery constant of ML-KEM is 2285, which is precisely 2 to the sixteenth power reduced modulo 3329, the Montgomery representation of unity. That the table begins with the representation of unity is the characteristic signature of Montgomery form, and it permits the generator's output to be verified against a known value before any binary is examined: the procedure is correct if and only if it reproduces these established table heads. The generator is accordingly the first component to construct and the most straightforward to test.

The observable that distinguishes implementations is not the representation adopted for a single table but the number of tables present, and this constitutes the basis of the attribution result of Section IX. The reference C lineage carries one forward table and reuses its second half, with runtime negation, for the base case, thereby storing 256 bytes of twiddles. The from-specification family, comprising the Go standard library and the RustCrypto crate, carries two plain tables, a forward table and a separate base-case table, totalling 512 bytes. The most widely deployed implementation, the one distributed in Chrome, carries three plain tables, adding an explicit inverse table, totalling 768 bytes. The total twiddle-byte count is therefore a single scalar that separates the three static lineages without any representation-level reasoning, and it survives every transformation that the tables themselves survive. Figure 3 presents the derivation and the mapping from table count to lineage.

The ML-DSA table is generated by the same procedure with $q = 8380417$, its own root, an eight-bit reversal, and 256 signed thirty-two-bit entries. One detail of the reference table warrants recording because it becomes relevant subsequently: the reference sets the unused zeroth entry to 0 rather than to the value that equation 1 would produce, so that an exact match against a first-principles table fails on that single entry even before a compiler is involved. The scoring must accommodate this, which, as Section VI argues, it is required to do in any event.

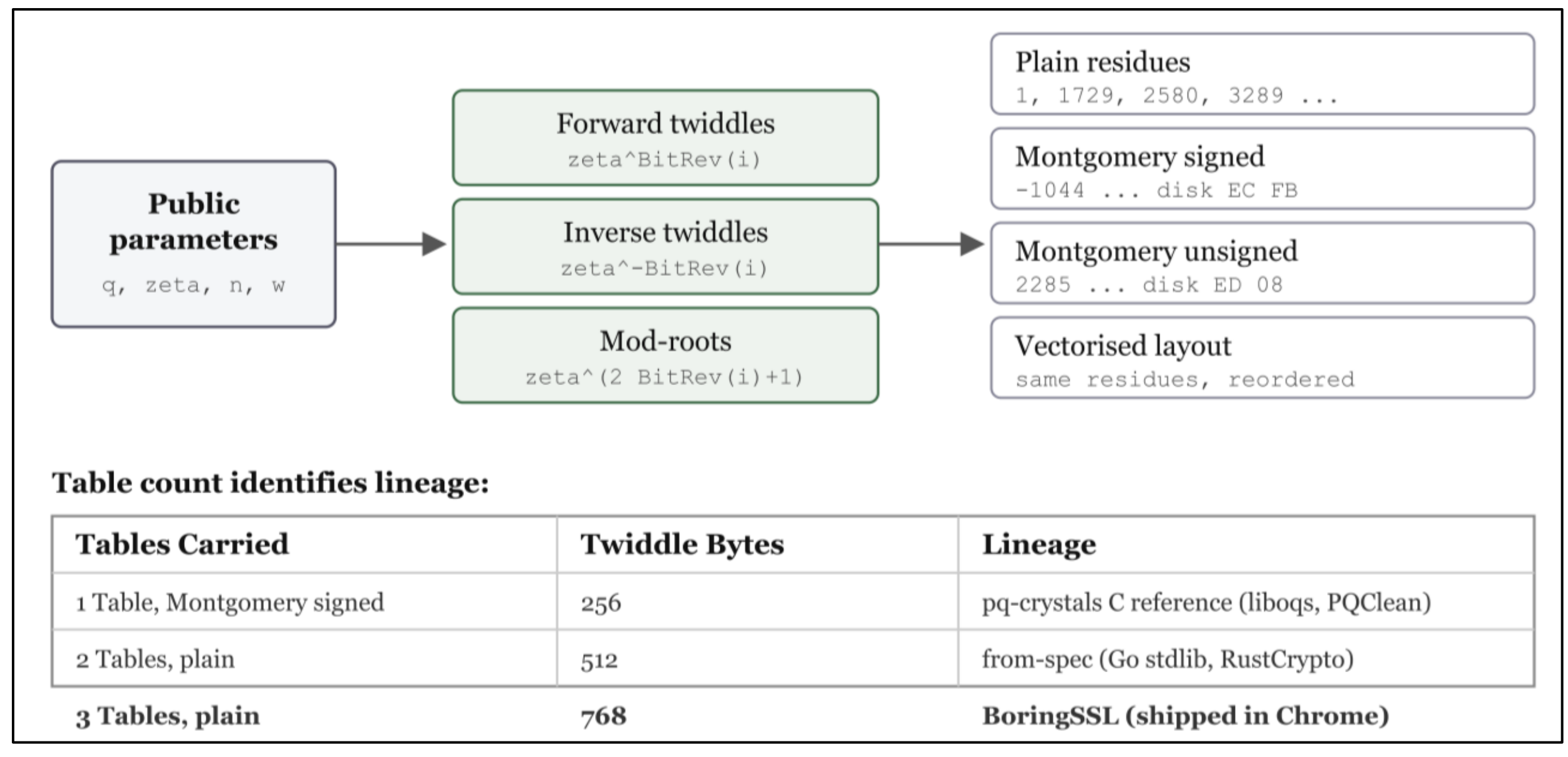

| Tables Carried | Twiddle Bytes | Lineage |
|---|---|---|
| 1 Table, Montgomery signed | 256 | pq-crystals C reference (liboqs, PQClean) |
| 2 Tables, plain | 512 | from-spec (Go stdlib, RustCrypto) |
| **3 Tables, plain** | **768** | **BoringSSL (shipped in Chrome)** |

*Fig. 3. One procedure derives every representation a real implementation stores, from public parameters alone*

## VI. Detector Design

The detector does not search for a byte string. A byte-level search fails for a reason that is structural rather than incidental: the same table exists in at least four byte-level forms, vectorised builds reorder it, and even the reference table diverges from a first-principles table on one entry. A byte matcher would require a distinct special case for each of these conditions. The design instead normalises and scores, which converts each of these obstacles into a parameter of a single mechanism. Figure 4 sets out the resulting pipeline.

### A. Normalisation

The detector advances a window across the read-only data section, reading sixteen-bit values for ML-KEM and thirty-two-bit values for ML-DSA. Each candidate value is normalised to a canonical residue modulo q. Signed values are folded to their unsigned equivalent, and Montgomery-form values are converted back to plain residues by multiplication by the inverse Montgomery factor, R inverse modulo q, which equals 169 for ML-KEM, this being evaluated as a second hypothesis alongside the raw residue. Following normalisation, the plain, signed-Montgomery, and unsigned-Montgomery representations of a given constant all reduce to the same canonical value, so that a single target set matches all three.

### B. Multiset scoring

A window is scored according to the number of expected canonical residues it contains, treated as a multiset and without regard to order. This order-independence is what accommodates the vectorised layout without a special case, since a reordered table contains the same residues and therefore attains the same score. The detector applies a threshold to the count of matched residues rather than to exact equality with a reference table. Threshold scoring is not a concession introduced for vectorised builds; it is required from the outset, because the reference ML-DSA table already diverges from a first-principles table on its zeroth entry, and a design demanding exact equality would fail against ground truth.

### C. False-positive bound

Because the target is a specific set of residues drawn from a space of q possible values, the probability that a window of unrelated data contains a large proportion of them by chance is small and, more usefully, computable. The likelihood that a random window contains one hundred or more distinct members of a specific 128-element subset of a 3329-element space is exceedingly small, and the paper reports this as an explicit quantity rather than as a general reassurance. The threshold is then set so as to place the false-positive rate below a stated bound. A refinement weights each matched residue according to its rarity in ordinary data, so that small values such as 1 or 17, which occur frequently in unrelated integer data, contribute little evidence, whereas large and distinctive residues contribute substantially; the score thereby becomes a sum of evidence expressed in bits, and the threshold a defensible likelihood rather than an arbitrary value.

### D. A negative filter, and a rejected signal

One further constant provides corroboration, and it must be handled with care in order to avoid the very failure this paper seeks to prevent. Every ML-KEM and ML-DSA implementation uses the SHAKE extendable-output function and therefore carries the twenty-four-entry round-constant table of the underlying Keccak permutation in its read-only data, where it survives precisely what the zeta table survives. Keccak constants are not a discriminator in isolation, since SHA-3 is ubiquitous. It would be tempting to employ their absence as a suppressing filter, discarding any zeta match not accompanied by Keccak constants. This design declines that approach, because silent suppression reproduces the same silent-failure behaviour that renders name-based tools hazardous: it destroys evidence that the analyst never observes. The detector instead reports the Keccak result as a condition presented alongside the score. A high zeta score accompanied by Keccak constants is reported as a strong, corroborated identification, whereas a high zeta score without them is reported as an identification with a noted absence of corroboration, which the analyst may weigh accordingly, since a binary may legitimately contain the lattice code while its SHA-3 implementation is supplied by a separately linked dependency. The scope of the corroboration check is therefore stated together with the result: it is applied within the static closure of the file and its statically bound dependencies. The reduction constant q-inverse, by contrast, is deliberately not employed as a primary signal, for reasons Section VIII-E demonstrates by measurement.

### E. Candidate signals evaluated and rejected

The reduction constant q-inverse and the fixed public-key and ciphertext sizes are superficially attractive fingerprints, and both are rejected here on principled rather than practical grounds, since rejecting them with a stated reason strengthens the method. The reduction constant is compiled into the instruction stream as an immediate operand, and the artefact sizes are typically compile-time buffer definitions; both therefore reside in the code section, the region that optimisation rewrites and obfuscation targets. The central argument of the paper is that data survives what code does not, and that argument applies to the zeta table but not to these values. They are consequently reported as corroborating signals with measured recall, never as primary discriminators. The evaluation of a method's own candidate signals, and the discarding of those that fail, confers greater credibility than their inclusion would.

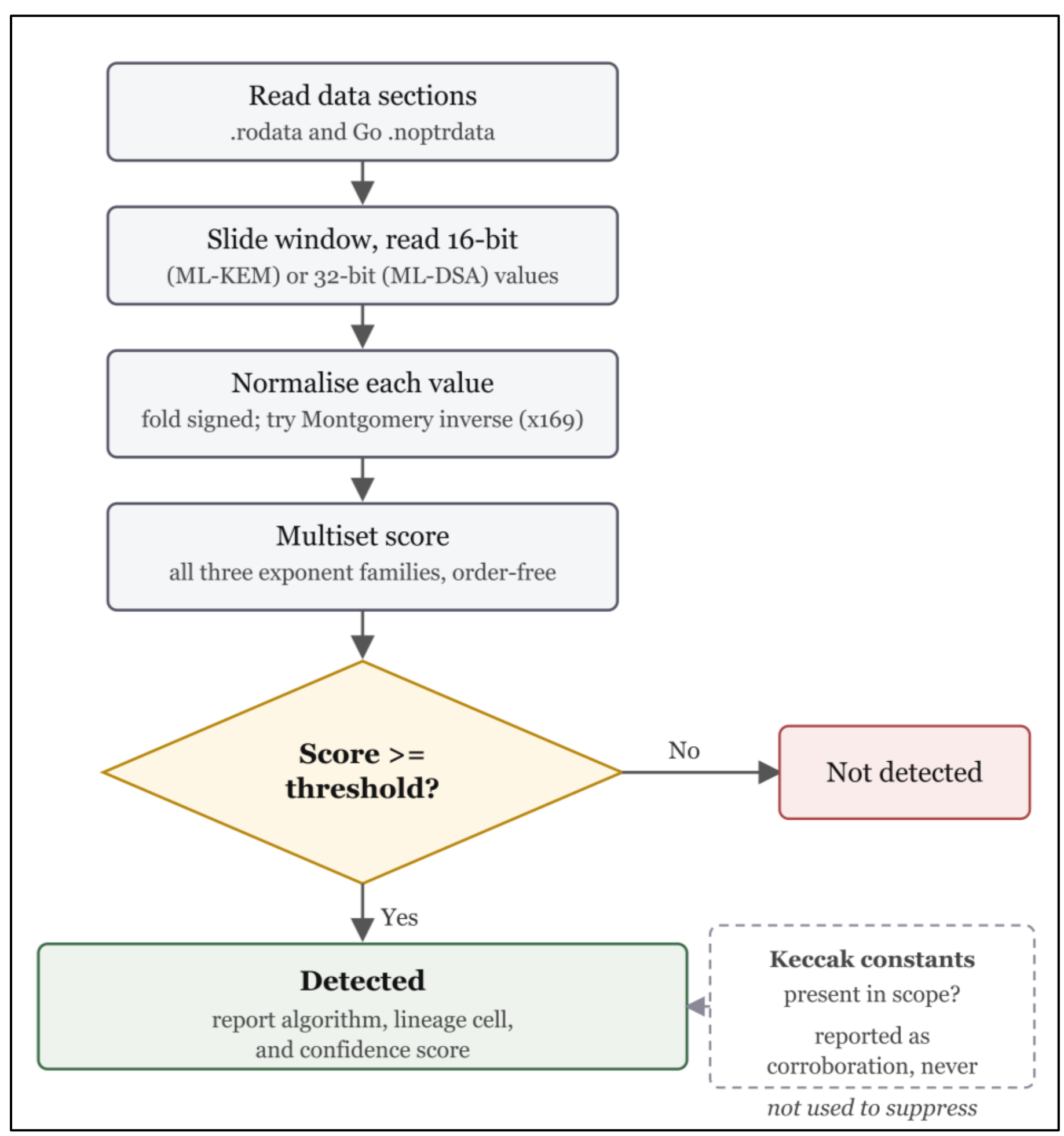

*Fig. 4. The detection pipeline normalises and scores rather than matching bytes. The Keccak check is attached to the result as a reported condition, not a gate that discards matches.*

## VII. Methodology and Implementation

This section describes the intended construction of the artefact [30] and the corpus on which it will be evaluated. It is written before the empirical work so that the design is fixed independently of its results, and the measured outcomes reported in Section VIII follow the protocol set out here without modification.

### A. Order of construction

The generator is constructed and tested first, in isolation, because every subsequent component depends upon it and because it is readily verifiable against the known table heads recorded in Section V. Only once the generator reproduces those constants exactly is a single ground-truth binary constructed and its read-only data extracted and inspected directly, in order to confirm that the table predicted from source and arithmetic in fact appears in a compiled artefact in the expected representation. This single confirmation converts the foundation of the method from a claim concerning source code into an observation concerning binaries, and it precedes all work on the detector. The detector, the scoring harness, and the evaluation matrix are constructed upon that confirmed foundation.

### B. Ground-truth corpus

The corpus is assembled from open implementations spanning the lineage cells the generator distinguishes: the reference C code and the liboqs and PQClean libraries derived from it [20], [21]; a vectorised build to exercise the reordered layout [17]; the Go standard library and the RustCrypto crate as the from-specification family; and the widely deployed implementation that ships in Chrome [24], which carries the three-table layout and is the most consequential single target because it is the most installed ML-KEM in existence. Each binary has a known answer, since its source is known, which makes the corpus a labelled test set. Negative samples are drawn from classical-only cryptographic binaries and from substantial non-cryptographic binaries, and a deliberately hard positive class is constructed from hybrid binaries that carry both a classical primitive and ML-KEM in one file, the arrangement that real deployments, and the observed ransomware, actually ship.

### C. Transformation matrix

Each source program is compiled into a matrix of variants that reproduces the transformations software undergoes between authoring and distribution: an unoptimised baseline retaining symbols, an optimised build, a stripped build, a statically linked build, and an obfuscated build. Applying the detector across this matrix measures the degree to which detection is maintained as the transformations accumulate. This matrix constitutes the validation of the detector rather than the principal claim of the paper. The paper does not contend that other tools fail; it measures the conditions under which its own method holds and those under which it degrades.

The matrix is not uniform across languages, and it is reported as a per-language equivalence rather than as an implied single grid, since a reader familiar with any particular toolchain would otherwise have reason to doubt the evaluation. The C column takes the expected form, comprising an optimisation flag, a separate stripping step, an explicit static-linking option, and a source-to-source obfuscator. The Go column has no equivalent optimisation flag, links statically by default, and retains its program-counter line table through the ordinary stripping step, so that a different obfuscator is required and the stripped and static cells signify something different from their counterparts in C. The Rust column optimises through its release profile rather than through an isolated flag. Each cell therefore states the concrete toolchain action it represents, and cells with no meaningful equivalent in a given language are marked accordingly rather than left blank or silently equated.

**TABLE I. Per-language transformation equivalence. The concrete build action behind each matrix column, per toolchain.**

| Transformation | C (gcc/clang) | Go | Rust |
|---|---|---|---|
| Baseline | -O0, symbols kept | go build, symbols kept | rustc default |
| Optimised | -O2, -O3, -Os | n/a (no -O levels) | rustc -O |
| Stripped | strip | -ldflags "-s -w" | strip |
| Static | -static | static by default | static by default |
| Obfuscated | LLVM instruction substitution | n/a | n/a |

### D. Environment and provenance

The entire evaluation runs on a commodity virtual server and requires no specialised or physical hardware, which keeps the work reproducible and within the resources of an independent researcher. Tooling versions, compiler flags, and library commits are recorded in full so that the corpus and its variants can be regenerated exactly, and the provenance of every corpus entry, including which lineage cell it belongs to, is given in Table II.

**TABLE II. Corpus provenance. Every implementation compiled, with toolchain and lineage cell, so the corpus is reproducible.**

| Implementation | Lang. | Version | Compiler / flags | Lineage cell | Twiddle bytes |
|---|---|---|---|---|---|
| pq-crystals Kyber | C | master (13 Aug 2026) | gcc 15.2 / clang 21, -DKYBER_K=4 -O2 | one-table C ref | 256 |
| pq-crystals Dilithium | C | master | gcc 15.2, -DDILITHIUM_MODE=5 -O2 | one-table C ref | 1024 |
| Go stdlib mlkem | Go | 1.26.5 | go build | two-table from-spec | 512 |
| RustCrypto ml-kem | Rust | 1.95.0 | rustc -O | two-table from-spec | 512 |
| BoringSSL | C | master | g++ -O2 | three-table BoringSSL | 768 |

# VIII. Evaluation

The evaluation comprises two parts. The first measures the detector on a labelled corpus, in which the correct result for every binary is known because its source was compiled directly. The second applies the detector to a stock system, in which the result is not known in advance, and it is the second that produced the central finding of the paper. Table III summarises each experiment and its principal outcome in advance of the detailed results that follow.

**TABLE III. Summary of experiments and results. Each row is expanded in the detailed tables and text that follow.**

| Experiment | What it tested | Headline result |
|---|---|---|
| Transformation matrix | recall under optimisation, stripping, static linking, obfuscation | full table match across four lineages, perfect precision |
| System sweep | false positives over 6,224 stock binaries | zero false positives; 24 real ML-KEM deployments found |
| Genuine vs bluff | real ML-KEM against a binary that only advertises it | genuine detected, bluff correctly rejected |
| Analytical bound | false-positive probability vs measurement | measured zero, consistent with the bound |
| Adversarial evasion | four attempts to defeat the detector | three cheap evasions failed; only byte-rewrite succeeded |
| Forensic recovery | detection from a disk image, including a deleted file | recovered from unallocated space after carving failed |
| Performance | scan cost and scale | full sweep in about ten minutes, no special hardware |

## A. Recall across the transformation matrix

Each corpus binary was compiled into every applicable variant and scanned. Precision was 100 per cent on every positive, so that only recall is reported. The result is that the table survives every transformation an ordinary build applies. The single consistent departure from a perfect score is the ML-DSA reference, which reads 255 of 256; this reflects the zeroed first table entry described in Section V and is present identically in every column rather than constituting a degradation caused by any transformation. Figure 5 plots this result against a name-based baseline, which collapses to zero once symbols are removed [31].

**TABLE IV. Recall across the transformation matrix, matched residues over table size. A cell marked n/a means the transformation has no equivalent in that toolchain (for example, Go exposes no optimisation levels and links statically by default), not a detection failure.**

| Lineage cell | Base | -O2 | -O3 | -Os | Strip | Static | Obf. |
|---|---|---|---|---|---|---|---|
| C ref, ML-KEM | 128/128 | 128/128 | 128/128 | 128/128 | 128/128 | 128/128 | 128/128 |
| C ref, ML-DSA | 255/256 | 255/256 | 255/256 | 255/256 | 255/256 | 255/256 | 255/256 |
| Go, ML-KEM | 128/128 | n/a | n/a | n/a | 128/128 | n/a | n/a |
| Rust, ML-KEM | 128/128 | 128/128 | n/a | n/a | 128/128 | n/a | n/a |
| BoringSSL, ML-KEM | 128/128 | 128/128 | 128/128 | 128/128 | 128/128 | n/a | n/a |

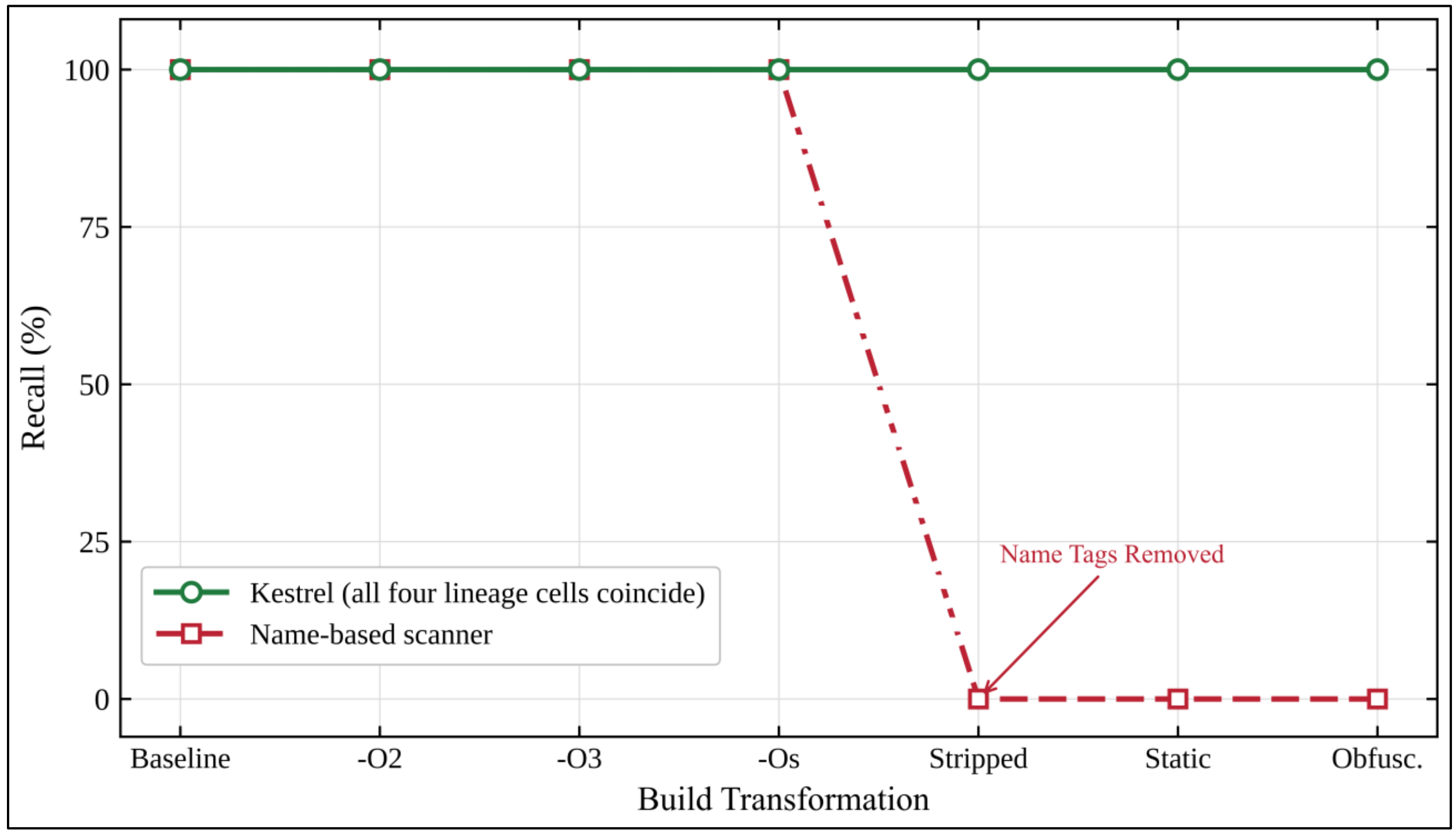

*Fig. 5. Recall across accumulated build transformations. All four lineage cells coincide at full recall and are drawn as a single line; a name-based scanner collapses to zero once symbols are stripped, while the constant-table signal is unaffected.*

### B. A false-positive sweep that became a discovery

The detector was applied to every binary in /usr/bin and /bin on a stock system, 6,224 files in total, in order to measure the frequency with which it responds to code containing no lattice cryptography. It produced no false positives. The noise floor, defined as the highest score attained by any binary without post-quantum cryptography, was 27 against a threshold of 32, while every genuine detection scored the full 128. The result is distinctly bimodal: noise remains below 27, signal occurs at 128, and the interval between them is empty. The threshold is not positioned at a boundary but within a wide unoccupied band, which constitutes the strongest evidence that the operating point was not selected so as to improve the apparent outcome [32].

The sweep was designed to identify false positives, and it instead disclosed genuine detections. Twenty-four of the 6,224 scanned paths genuinely contain ML-KEM, none of them catalogued anywhere as doing so; because /bin is symlinked to /usr/bin on this system, these resolve to twelve distinct programs, each detected on both paths. They fall into two groups, and the two groups correspond to two different lineages, which represents the attribution result appearing without prompting on production software the tool had not previously encountered.

**TABLE V. False-positive sweep over a stock system, and the genuine post-quantum crypto it uncovered.**

| Category | Binaries | False positives | Top score on a true negative |
|---|---|---|---|
| /usr/bin + /bin, full sweep | 6,224 | 0 | 27/128 |
| of which genuine ML-KEM | 24 paths (12 distinct) | (true positives) | 128/128 each |

OpenSSH on the test system offers mlkem768x25519-sha256 as a hybrid post-quantum key exchange, confirmed by direct interrogation of its supported algorithms, so that this constitutes an active key-exchange capability rather than an incidental table within a linked library. The algorithm was introduced in OpenSSH 9.9 and established as the default for key agreement in OpenSSH 10.0, released in April 2025, so that a current system negotiates it without additional configuration. The detector located the corresponding ML-KEM table within the stripped ssh binary. The fingerprint was of Montgomery form, single-table, and situated in the .rodata section, consistent with a C-library implementation such as OpenSSL rather than with a from-specification port. OpenSSH negotiates ML-KEM-768 in this instance rather than the ML-KEM-1024 of the constructed corpus; the two parameter sets share the same zeta table, since the constants vary only with the ring and not with the security level, so that the detection is valid for either, and the parameter set is stated precisely in order to correspond to what the binary actually offers [41].

The container tooling constitutes the second group. Across containerd, docker, runc, and ctr the detector identified one identical fingerprint: two plain-residue tables in Go's .noptrdata section. This is the ML-KEM implementation of the Go standard library, introduced through the toolchain rather than through a deliberate decision by the authors of each project. Post-quantum primitives are thus propagating into production binaries without the awareness of the teams distributing them, which is precisely the visibility gap that a cryptographic bill of materials is intended to close, demonstrated here on deployed software rather than argued in principle. Cryptography that cannot be observed cannot be inventoried, and this cryptography was, until now, largely unobserved, residing in stripped binaries that no current inventory tool identifies [42].

The two groups additionally demonstrate the attribution contribution on previously unseen binaries. The ssh binary was identified as C-reference lineage, Montgomery form, one table, in .rodata. The container binaries were identified as from-specification lineage, plain form, two tables, in .noptrdata. The detector distinguished an OpenSSL-linked implementation from a Go-runtime implementation by table configuration alone, on production software that neither the tool nor its author had previously examined, which establishes the attribution claim in concrete rather than asserted terms.

### C. The analytical bound

The measured value of zero is explained rather than merely observed. Modelling the contents of a window as uniform-random values over the 3,329 residues of the ring yields a per-value probability of approximately 0.0385 of falling within the 128-member target set, and the number of chance matches within a window follows a binomial distribution whose upper tail above the threshold constitutes the per-window false-positive probability. The table below reports that probability, the expected false-positive count over the 6,224-binary sweep, and the corresponding count over an 8 GB image, at several thresholds [33].

**TABLE VI. Analytical false-positive bound, uniform-random model, as a conservative ceiling. Expected counts are given for the 6,224-binary sweep and for an 8 GB image.**

| Threshold | Per-window bound | Expected FP (6,224 sweep / 8 GB image) |
|---|---|---|
| 32 | 6.4e-9 | ~4 / ~27 |
| 40 | 6.9e-14 | ~0 / ~0 |
| 48 | 1.2e-19 | ~0 / ~0 |

| Threshold | Per-window bound | Expected FP (6,224 sweep / 8 GB image) |
|---|---|---|
| 64 | 3.2e-33 | ~0 / ~0 |

The bound is a conservative ceiling. It assumes uniform-random data and does not encode the requirement that the matched residues be distinct, and it therefore overstates the true rate. Applied specifically to the 6,224-binary sweep, the same model predicts approximately four expected false positives at threshold 32, and approximately twenty-seven for an 8 GB image, whereas the measured count was zero in both cases: structured data does not behave as the uniform worst case, and the distinctness requirement that the model omits suppresses chance matches substantially. Raising the threshold to 40 reduces even the uniform worst case below ten to the negative thirteenth, while every genuine detection continues to score the full 128, so that the operating point at 32 lies within a wide safety margin rather than at a finely tuned boundary. Because real data departs from the model, the measured rate constitutes the primary evidence and the analytical bound the supporting argument, rather than the reverse.

### D. Genuine versus bluff

A binary can claim post-quantum cryptography it does not have. The detector reports what is present, not what is advertised, which is a determination no dynamic or network method can make from a file that is never run [34].

**TABLE VII. Genuine-versus-bluff discrimination. The bluff binary advertises ML-KEM in its strings while using classical cryptography.**

| Binary | Claims PQC | Contains PQC | Zeta table | Verdict |
|---|---|---|---|---|
| Real ML-KEM-1024 | no | yes | present | DETECTED 128/128 |
| Bluff (strings only) | yes | no | absent | NOT DETECTED 0/128 |

The bluff binary advertised the phrase ML-KEM-1024 post-quantum encryption within its strings while performing classical cryptography, and it scored zero. A string-based or name-based scanner reads the label and accepts it, whereas the table detector observes that no table is present. A live ransomware sample was deliberately not tested, since the acquisition of active ransomware is not warranted for this result and the bluff case already demonstrates the same capability, namely the safe identification of a false post-quantum claim.

### E. Rejected signals, and why

Two candidate signals were evaluated and relegated to a corroborating role, and the measurements demonstrate the reason. The reduction constant QINV and the fixed public-key size reside in the code section rather than the data section, so that they do not survive the transformations the zeta table survives, and where they do occur they occur by chance.

**TABLE VIII. Rejected corroborating signals. Match counts by section show these constants do not behave like the zeta table.**

| Signal | Section | Baseline | Stripped | Static |
|---|---|---|---|---|
| QINV | data (.rodata) | 0 | 0 | 0 |
| QINV | code (.text) | 1 | 1 | 7 |
| pk-size 1568 | data (.rodata) | 0 | 0 | 10 |
| pk-size 1568 | code (.text) | 6 | 6 | 13 |

QINV never appears in the data section in any build, which confirms that it is a code-section immediate operand. The ten data-section matches for the key size in the static build are coincidental byte collisions within the volume of statically linked libc, a result that supports rather than undermines the argument: a two-byte constant collides by chance and is therefore of no value as a fingerprint, for the same reason that the modulus 3329 alone carries no evidential weight. Only the ordered table provides such evidence [35].

### F. Adversarial evasion and the method's boundary

The detector was attacked directly, with each evasion's cost recorded, because the cost is what separates a trivial evasion that fails from an expensive one that works [36].

**TABLE IX. Adversarial evasion. Cheap evasions leave the residues in the file and fail; only transforming the stored bytes succeeds, and it costs real work.**

| Evasion | Effect on table | Verdict | Effort |
| --- | --- | --- | --- |
| Reorder table | residues present, shuffled | DETECTED 128/128 | trivial |
| Padding between entries | residues present, spaced | DETECTED 86/128 | trivial |
| Split across two sections | residues present, split | DETECTED 64/128 | easy |
| XOR with runtime undo | stored bytes transformed | NOT DETECTED 8/128 | moderate |

The three low-cost evasions all fail, since each leaves the genuine residues within the file and the multiset scorer identifies them irrespective of order, spacing, or section boundary. The scores decline from 128 to 86 to 64 but never cross the threshold. Only the XOR transformation defeats the detector, because it is the sole evasion that alters the stored bytes so that the residues are no longer present, and it requires substantive modification of the cryptographic code.

The same boundary is apparent from the opposite direction. A runtime-generated ML-KEM binary, which computes the twiddle table at load time and demonstrably stores nothing in the data section, scored 8 of 128 and was not detected, while nonetheless performing a correct key exchange. Static constant fingerprinting cannot identify an implementation that stores no constants. This is the genuine limit of the method; it was predicted in writing before the binary was constructed, and the prediction was borne out. Each of the twelve programs disclosed by the sweep had followed the conventional practice of storing its tables, which is the reason they were detectable [37].

### G. Performance

Kestrel advances a window across a data section and is therefore computationally inexpensive, and the measured cost confirms that the forensic and audit applications are practical at scale. The figures below were obtained on a Linux virtual machine allocated four processor cores and 8 GB of memory, hosted on an Intel Core i9-14900HX system, each scan being the median of five runs following a discarded warm-up, with the tool unmodified from its evaluation form.

**TABLE X. Detection performance across binary sizes. Throughput rises with size as the fixed per-invocation cost amortises.**

| Binary | Size (B) | Sections | Median scan (ms) | Throughput (MB/s) | Peak RSS (KB) |
| --- | --- | --- | --- | --- | --- |
| ML-KEM corpus | 36,976 | 3 | 33 | 1.1 | 14,464 |
| ssh | 1,076,256 | 4 | 169 | 6.4 | 27,000 |
| static binary | 786,152 | 4 | 109 | 7.2 | 27,588 |
| containerd (Go) | 40,951,704 | 4 | 3,917 | 10.5 | 395,664 |

The complete sweep of 6,224 binaries was performed in 583.6 seconds, approximately ten minutes, at a mean of 93.8 ms per binary over 2,704 MB of code. Throughput on the smallest binary is modest, at 1.1 MB/s, because a fixed initialisation cost predominates when the file is small; this cost is amortised as size increases, and the 41 MB Go binary attains 10.5 MB/s. Peak resident memory scales with the largest binary scanned rather than remaining constant, since the tool reads a file and its sections into memory, which is why the 41 MB binary raises the figure to 395 MB. For the stated scenarios, namely an examiner scanning a device or an auditor scanning a fleet, a duration of ten minutes for a complete system directory is well within practical limits. [38]

### H. Detection on recovered evidence

The scenario for which the paper is intended is that of a seized device rather than a live filesystem, and the final experiment accordingly placed Kestrel within a recovery workflow. A 200 MB ext4 image was constructed, three known binaries were copied into it, one was deleted, and the files were recovered using The Sleuth Kit [25] and foremost, the standard forensic tools. The detector was applied, unmodified, to whatever each recovery procedure returned.

**TABLE XI. Detection on binaries recovered from a 200 MB ext4 disk image. Live and carved rows match on verdict, score, and lineage.**

| Binary | Source condition | Recovery tool | Recovered | Verdict | Score | Lineage |
|---|---|---|---|---|---|---|
| ssh | live | none | yes | detected | 128/128 | C-reference |
| ssh | carved | Sleuth Kit icat | byte-perfect | detected | 128/128 | C-reference |
| containerd | live | none | yes | detected | 128/128 | from-spec |
| containerd | carved | Sleuth Kit icat | byte-perfect | detected | 128/128 | from-spec |
| corpus ML-KEM | live | none | yes | detected | 128/128 | C-reference |
| corpus ML-KEM | deleted, carved | Sleuth Kit / foremost | failed (0 B) | not detected | 0/128 | n/a |
| raw image | unallocated scan | Kestrel direct | table intact | detected | 128/128 | from-spec |

The two files that were recovered returned byte-perfect and were detected identically to their live originals, agreeing on both score and lineage, which is the comparison the experiment was designed to establish. Recovery alters nothing the detector reads, since a binary on disk consists of the same bytes however it is obtained [39].

The deleted file was not recovered, and the reason is a property of the filesystem rather than of the tool. Modern ext4 zeroes the block pointers of the orphaned inode upon deletion, so that Sleuth Kit returned zero bytes, and the ELF header was destroyed sufficiently that foremost's signature carver identified nothing. Conventional recovery depends upon file structure, which in this case had been destroyed.

That failure produced the most significant result in the paper. Although no tool could reconstruct the deleted file, its data section, including the table, remained present in unallocated space. Scanning the raw 200 MB image directly as a flat byte stream, without any file recovery, Kestrel located the ML-KEM table at offset 0x8c6420 and scored 128 of 128. Because detection depends upon constant tables rather than upon file structure, it survives precisely the file-level destruction that defeats a carver, and it operates upon a raw image as readily as upon intact files. One qualification concerns attribution: a raw scan of a mixed image reports the predominant lineage present, so that per-file attribution still requires per-file scanning, which is the step an examiner performs following carving in any case. The capability relevant to triage, namely confirming that post-quantum cryptography is present somewhere on a device whose files are damaged or deleted, holds on the raw image alone [40].

# IX. Implementation-Lineage Attribution

Detection determines whether a post-quantum primitive is present. A second question, and for an examiner a more valuable one, concerns which implementation is present, and the lineage cells of Section V address it at the granularity of a source lineage. The inventory of tables a binary carries is a property the implementer selected for reasons of arithmetic and performance rather than concealment, and it cannot be altered without rewriting the transform, which renders it a durable provenance marker that survives stripping and static linking together with the tables themselves.

Direct inspection of the implementations establishes the taxonomy independently of the corpus, which is why it is presented as a finding rather than as a hypothesis. The reference C lineage stores a single Montgomery-form table and reuses its second half for the base case, and its ML-DSA table carries a zeroed zeroth entry where a from-specification derivation would carry the true value; that zero is itself a provenance marker indicating reference-derived code. The from-specification family stores two plain tables, and the most widely deployed implementation, that in Chrome, stores three. These are separated by the single scalar of total twiddle-byte count, 256 as against 512 as against 768, prior to any representation-level reasoning. Section VIII-B demonstrates this on production software the tool had not previously encountered: the stripped ssh binary was identified as C-reference lineage from a single Montgomery table in .rodata, while the container tooling was identified as from-specification lineage from two plain tables in Go's .noptrdata, so that the detector separated an OpenSSL-linked implementation from a Go-runtime implementation by table configuration alone.

The principled limit of this contribution is that attribution is to a lineage rather than to a specific library. The Go standard library and the RustCrypto crate were both written from the specification and carry the same two plain-residue tables, identical in content; they store them in different sections, .noptrdata for the Go runtime and .rodata for the Rust build, but the constants themselves are indistinguishable, and separating the two libraries requires language-level metadata that lies outside the fingerprint. The paper claims lineage-level attribution, which is what an examiner can defensibly assert, and states explicitly that library-level attribution within the from-specification family is not obtainable from the constants alone. That two independent implementations in different languages converge upon the same tables is not a weakness of the method; it is evidence that the constants are a property of the standard rather than of any individual codebase, which is the same fact that lets the detector generalise to implementations it has not previously encountered.

# X. Limitations

The boundaries of the method are stated explicitly, since a limitation identified by the author constitutes a statement of scope, whereas the same limitation identified by a reviewer constitutes a deficiency. Five are of consequence.

**Runtime generation:** An implementation that computes the transform constants during initialisation rather than storing them presents no table to detect. This is the fundamental evasion; it is inexpensive, and it is available to any informed reader. It is, however, absent from every implementation examined and from the malware observed to date, since there is no incentive to conceal a primitive from a tool that did not previously exist. The method addresses cryptography obscured by ordinary build practice rather than cryptography deliberately concealed by an informed adversary, and it claims nothing further.

**Scheme coverage:** The approach covers lattice schemes that store a transform table. Hash-based SLH-DSA possesses no such table and lies outside the scope; the FFT-based signature scheme employs floating-point constants that present a distinct problem; and code-based schemes carry different structures again. Detectability is a function of a scheme's algebraic structure, and hash-based constructions are the most resistant precisely because they are built from primitives already in widespread use.

**Managed runtimes:** The method assumes native compilation, in which an array initialiser becomes a contiguous constant region. A bytecode or managed-runtime target may emit the same constants as a sequence of store instructions rather than as a data table, which a data-section scanner would not observe; their identification in that setting requires a bytecode-aware extractor. The fingerprint set is unchanged, whereas the extractor is not, and this is recorded as future work rather than claimed as covered.

**Dependency scope:** The Keccak corroboration check assumes that the permutation's constants and the zeta table reside within the same analysis unit. A binary that statically links its lattice code but dynamically links its SHA-3 implementation could carry the zeta table without the Keccak constants, so the check is applied within the static closure of a file and its statically bound dependencies, and its scope is reported rather than assumed.

**Adversarial table modification:** An adversary with write access to a binary can alter the stored constant tables directly, by permuting, encoding, or corrupting the residues, so that the detector no longer matches them, at the cost of adding code that reconstructs the correct tables before the algorithm runs. Section VIII shows that low-cost modifications which leave the residues in the file are still detected, and that only a transformation removing the residues from the stored bytes evades the method. An adversary willing to modify and re-test the cryptographic code can therefore defeat static detection, as with any signature-based method. Detecting deliberate table obfuscation, for example by recognising the reconstruction code itself, is left to future work.

## XI. Discussion

The method is situated at the intersection of two developments that emerged together in 2026. In the forensic context, post-quantum cryptography became an offensive capability in deployed ransomware, and the discrimination of genuine from fraudulent implementations became a concrete requirement of examination rather than a hypothetical concern. In the assurance context, a federal executive order established legally binding dates for post-quantum migration, and the question of how a compliance claim is to be verified against a shipped binary acquired substantial significance. Both developments concern the same deficiency, and the same static, constant-based method addresses both, since each reduces to the identification of a primitive within a static file.

The relationship to the established signature-matching tradition of the past two decades warrants precise statement, since it constitutes both the foundation of the method and a potential line of criticism. The technique of matching fixed constants is long established, and its evadability is well understood. The tools founded upon it remain in general use, not despite that evadability but because the great majority of software has no reason to evade detection. The AES substitution box has been trivially removable for two decades and remains worth detecting. The same reasoning transfers directly to the post-quantum setting: the evasion is straightforward and largely theoretical, and the detection is valuable precisely because real implementations, including malicious ones, incorporate a library rather than implement a lattice transform independently. The method is effective by virtue of the adversary's reliance on convenience, and it states this openly.

A concluding observation generalises the contribution without overstating it. A cryptographic scheme is detectable by this method to the degree that its correctness depends upon stored, invariant constants. Lattice schemes are detectable because the transform requires a fixed table; hash-based schemes resist detection because they reuse primitives that carry no distinctive constant of their own. Detectability is therefore a property of the underlying mathematics, and the boundary of this method corresponds to the boundary of that property.

## XII. Conclusion

Cryptographic discovery has, to date, been oriented in a single direction, toward the quantum-vulnerable algorithms that a migration must remove, and has left the converse question unaddressed: the confirmation, from a compiled binary, that a quantum-safe algorithm is genuinely present. This paper addresses that question [30] for the lattice schemes that predominate in deployment, using the number-theoretic transform constants that ML-KEM and ML-DSA cannot compute without and that survive the compilation, stripping, and static linking which eliminate every name-based signal. The contribution comprises a derivation procedure that generates the complete fingerprint set from public parameters, a detector that normalises and scores rather than matching bytes and thereby tolerates the representations that real implementations employ, and a treatment of implementation-lineage attribution that furnishes an examiner with provenance rather than mere presence. The method is bounded candidly: it covers lattice schemes with stored tables, it addresses ordinary build practice rather than deliberate concealment, and it arrives at the moment at which post-quantum primitives have appeared in malware and a federally mandated migration deadline has rendered their verification a matter of compliance. The underlying technique is long established; this is its first application to the problem of confirming post-quantum deployment.

## Declaration of Competing Interest

The author declares that he has no known competing financial interests or personal relationships that could have appeared to influence the work reported in this paper. This research received no specific grant from any funding agency in the public, commercial, or not-for-profit sectors.

## Data Availability

Kestrel, comprising the fingerprint generator, the detector, the corpus-construction scripts, and the LLVM obfuscation pass used in the transformation matrix, together with every benchmark log, corpus script, and detector output reported in this paper, is openly available in the public repository at https://github.com/mshaheerjunaid/Kestrel, so that every result reported here can be independently reproduced. The large BoringSSL and obfuscated binaries are regenerated by the included build scripts rather than distributed directly. Specific artefacts are cited inline against the results they support, and the exact software environment on which all measurements were taken is recorded alongside them [30].